\documentclass[10pt,letterpaper]{article}
\usepackage{opex3}
\usepackage{graphicx}
\usepackage{amsmath,amsthm,amssymb}

\begin{document}

%
%

\title{Broadband Extraordinary Transmission in a Single Sub-wavelength Aperture}

\author{Wenxuan Tang,$^1$ Yang Hao,$^{1,*}$ and Francisco Medina $^2$}

\address{$^1$Department of Electronic Engineering, Queen Mary University of
London,\\ Mile End Road, London E1 4NS, United Kingdom
\\
$^2$Department Electronics and Electromagnetism, Physics Faculty,
University of Seville, \\ Av. Reina Mercedes s/n, 41012-Seville,
Spain}

\email{$^*$yang.hao@elec.qmul.ac.uk} 



\begin{abstract} Coordinate transformation is applied to design an all-dielectric device for
Extraordinary Transmission (ET) in a single sub-wavelength slit. The
proposed device has a broadband feature and can be applied from
microwave to visible frequency bands. Finite-Difference Time-Domain
(FDTD) simulations are used to verify the device's performance. The
results show that significantly increased transmission is achieved
through the sub-wavelength aperture from 4 GHz to 8 GHz when the
device is applied. In contrast with previously reported systems, the
frequency sensitivity of the new device is very low.
\end{abstract}



\section{Introduction}
The extraordinary (optical) transmission (EOT) of electromagnetic
waves through periodic arrays of subwavelength holes
\cite{Ebbesen98} or slits \cite{Porto99} made in opaque --- mostly
metallic --- screens has been exhaustively studied along the last
decade. Comprehensive review papers reporting on the basic physics
behind the phenomenon are available nowadays
\cite{Genet07,Garciadeabajo07,Garciavidal10}. The research of
periodic structures has also stimulated the study of the
transmission properties through single holes or slits. Sometimes the
physics of EOT through periodic structures is closely related with
the physics of enhanced transmission through single apertures, as it
is the case of slits or holes around which the conducting surface is
periodically structured \cite{Lezec02,Garciavidal03,Dunbar09}. The
periodically perforated screen or the periodically structured
surface (a metal surface with corrugations, for instance) support
surface waves (the so-called \textit{spoof plasmons}
\cite{Pendry04}) which are strongly excited around certain critical
frequencies. However, any other mechanism provoking field
enhancement at the aperture level would automatically induce
enhanced transmission. Thus, it seems almost obvious that high
quality factor resonators placed close to or inside the holes will
produce resonant enhancement of the fields and, consequently,
enhanced transmission \cite{Aydin09,Cakmak09}. Other methods have
been proposed to enhance transmission of electromagnetic waves
through electrically small slits or holes
\cite{Bilotti09,Valdivia10}.

A common feature to all the methods mentioned in the previous
paragraph is their intrinsic narrow band behavior. Narrow bandwidth
is associated with the resonant nature of the underlying cause of
the phenomenon. A completely different approach is proposed in this
contribution. The new approach to obtain extraordinary transmission
is based on light squeezing controlled by non-uniform and
non-resonant dielectric permittivity distributions around the
aperture. Coordinate transformation is applied to design the
permittivity distributions and appropriate approximations are
introduced to realize a practical all-dielectric device. This design
is tested by in-house FDTD based simulations and a great enhancement
is verified when the novel devices are applied to the sub-wavelength
aperture. Although some frequency dependence of transmissivity is
observed, the bandwidth of the relevant enhanced transmission
frequency region is much larger than any previously reported.

\section{The scheme to enhance the transmission}

The method of coordinate transformation offers us an easy way to
control the travelling route of the electromagnetic waves
\cite{Pendry06,Leonhardt06}. In a Cartesian coordinates, an incident
electromagnetic wave travels straightly, while in a distorted
coordinates it turns according to the mesh. On the other hand,
Maxwell's equations are form-invariant under coordinate
transformations \cite{Pendry06}, and, therefore, one can manipulate
the electromagnetic wave in a distorted space by engineering the
background with the necessary material parameters.
\begin{figure}[t]
\centerline{
\includegraphics[width=7.5cm]{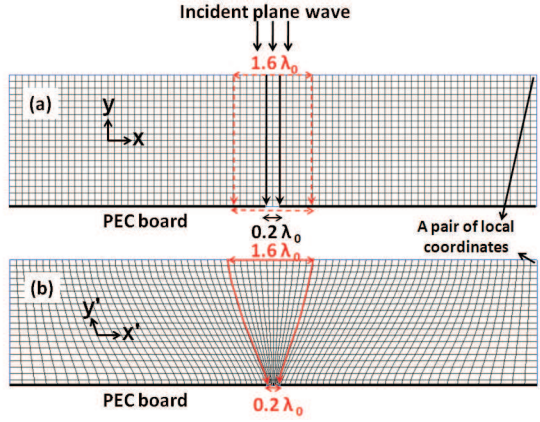}} \caption{(a)
The space in the Cartesian coordinates. (b) The space in the
distorted coordinates.} \label{label:2spaces}
\end{figure}

%

The main scheme to enhance the transmission through a sub-wavelength
aperture is to build a distorted space which can smoothly induce
much more energy from a larger aperture into a smaller one.
Fig.~\ref{label:2spaces}(a) shows a space in the Cartesian
coordinates. When a plane wave travels into the space from the top,
the wave vector $k$ is in the direction towards the bottom along the
edge of a grid, as illustrated with black arrows, and so is the
Poynting vector $S$, which represents the flowing of energy. For a
perfect electric conductor (PEC) with a $0.2\lambda_{0}$
($\lambda_{0}$ the wavelength of free space) wide slit, most
incident energy is reflected and little energy could pass through
the slit. When the aperture of the slit is enlarged to, for
instance, $1.6\lambda_{0}$, the transmitted energy increases
significantly. A distorted space is designed in
Fig.~\ref{label:2spaces}(b), where the wave vector $k$ is bending
towards the slit along the curvature of the grids and the energy is
guided from a virtual electrically-large aperture to an
electrically-small slit. In this way, the transmission through a
sub-wavelength slit could be enhanced dramatically in the distorted
space.

A discrete coordinate transformation is used to design the
background material for the distorted space. Instead of using the
general coordinate transformation in the whole space, the
transformation is operated between every pair of local coordinates
\cite{li2008hiding} (a pair of local coordinates is defined as the
example in Fig.~\ref{label:2spaces}). To simplify the problem, here
we only discuss the two dimensional (2D) case with $H_{x}$, $H_{y}$
and $E_{z}$ components. The Jacobian transformation matrix $J$ is
used to map one space to the other. The relative permittivity and
permeability values of each block in the distorted space are
calculated as \cite{Schurig}

\begin{equation}\label{eq:epsandmu}
\bar{\bar{\varepsilon'}}=\frac{J
\bar{\bar{\varepsilon}}J^{T}}{det(J)}, \bar{\bar{\mu'}}=\frac{J
\bar{\bar{\mu}}J^{T}}{det(J)}
\end{equation}
Notice that $J$ is defined for every pair of local coordinates and
it has a simpler form in this 2-D case as
\begin{equation}\label{eq:Jacibian}
J= \left(
\begin{array}{rrr}
\frac{\partial x'}{\partial x} & \frac{\partial x'}{\partial y} & 0 \\
\frac{\partial y'}{\partial x} & \frac{\partial y'}{\partial y} & 0 \\
0 & 0 & 1
\end{array}
\right)
\end{equation}

A practical way to realize the distorted space is to use an
all-dielectric background material. It has been proved that if all
the local coordinates in the distorted space are quasi-conformal,
the background permeability could be considered as unity and
quasi-isotropic for the E-polarization \cite{li2008hiding}, then,
only the permittivity map is required in the z direction. In this
scenario, it is difficult to generate a strictly-orthogonal grid to
map the distorted space linking the sub-wavelength slit with an
electrically large aperture. Instead, very fine meshes are carefully
generated in the distorted space to ensure most cells nearly
orthogonal, especially in the central region above the slit.
Although such an engineering approximation may not represent a
mathematically strict transformation, and, hence, is sensitive to
wave incident angles for the proposed device, the performance
enhancement in terms of extraordinary wave transmission is still
evident over a broadband of frequencies, as demonstrated in the
following sections.

\begin{figure}[t]
\centering
\includegraphics[width=7.5cm]{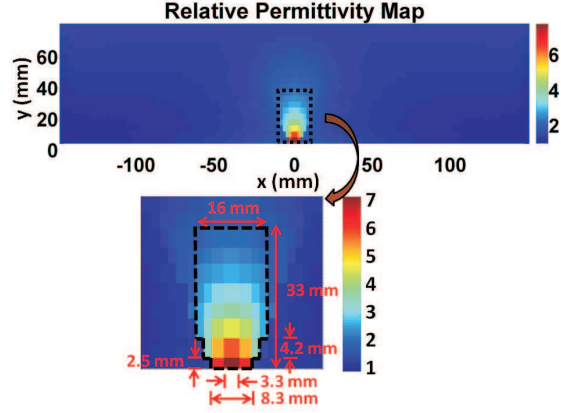}
\caption{The relative permittivity map of the distorted space. The
dashed black line outlines the profile of the enhancement device.}
\label{label:eps}
\end{figure}

The relative permittivity ($\varepsilon_{r}$) map is given in
Fig.~\ref{label:eps}. Relative permittivity is very close to unity
in most parts of the distorted space, which means most of the
background material could be considered as air and dielectrics with
varying $\varepsilon_{r}$ are mainly required in the central area
surrounding the slit. The dashed black line in Fig.~\ref{label:eps}
outlines the profile of such an enhancement device and the detailed
permittivity map. The device has a size of
$0.43\lambda_{0}\times0.88\lambda_{0}$ at 8 GHz.

\section{Enhancement performance at a single frequency}
\begin{figure}[hb]
\centering
\includegraphics[width=8cm]{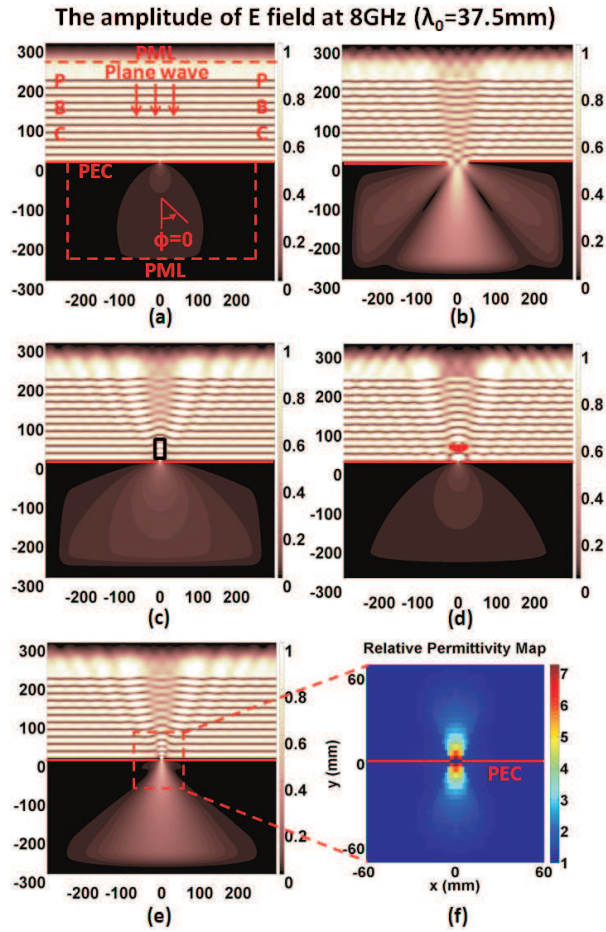}
\caption{ The amplitudes of the electric field in different cases.
(a) The incident plane wave illuminates a PEC plate with a
$0.2\lambda_{0}$ wide slit. (b) The incident plane wave illuminates
a PEC plate with a $1.6\lambda_{0}$ wide slit. (c) The incident
plane wave illuminates the above mentioned sub-wavelength slit when
the proposed enhancement device is applied in the black lines. (d)
The incident plane wave illuminates the sub-wavelength slit after
crossing a focusing lens with its focal point located on the slit.
(e) The incident plane wave illuminates a PEC plate with a pair of
enhancement devices at both sides of the sub-wavelength slit. (f)
The permittivity map around the slit in case (e). The map is
symmetric.} \label{label:field}
\end{figure}

FDTD simulations \cite{Yan,argyropoulosieeetrans} are used to test
the performance of the designed device without loss of generality.
The operating frequency is set to 8 GHz, the wavelength in free
space ($\lambda_{0}$) being 37.5 mm accordingly. The sub-wavelength
slit on the PEC plate is $0.2\lambda_{0}$ (7.5 mm) wide and
$0.067\lambda_{0}$ (2.5 mm) thick. The simulation domain is shown in
Fig.~\ref{label:field}(a). Periodic boundary conditions (PBC) are
applied to realize an incident plane wave with E-polarization
impinging from the top. Perfectly matched layers (PML) are added at
the left and right sides to eliminate the interference from the
neighbours \cite{Taflove05}. The amplitude of the electric field is
plotted to quantify the distribution of energy. In
Fig.~\ref{label:field}(a), the incident plane wave irradiates on the
PEC plate with the sub-wavelength slit. Extremely low field
intensity is observed on the other side of the plate.
Fig.~\ref{label:field}(b) shows the field distribution when the slit
is enlarged to be $1.6\lambda_{0}$ (60 mm) wide. Much energy passes
through the slit and propagates further, as expected. When the
proposed enhancement device, designed in Fig.~\ref{label:eps}, is
located over the $0.2\lambda_{0}$ wide slit (inside the region
defined by the black lines in Fig.~\ref{label:field}(c)), a
significant enhancement of the field is observed on the other side
of the PEC plate. Ideally, from the point of view of coordinate
transformation, the transmitted energy should be the same in (b) and
(c). However, the device includes only the central part of the
distorted space and the orthogonality in the distorted space is not
guaranteed everywhere. The anisotropic property needs to be included
for even enhancing the transmission. Nevertheless, the proposed
device still produces strong field enhancement, when comparing
figures ~\ref{label:field}(a) and ~\ref{label:field}(c). A
conventional focusing lens is also applied in
Fig.~\ref{label:field}(d) to compare the performances. The lens has
an aperture of $1.6\lambda_{0}$ and its focal point is set to be at
the $0.2\lambda_{0}$ wide slit to focus more energy directly inside
the slit. The transmission through the $0.2\lambda_{0}$ wide slit
increases only slightly. Thus the proposed device outperforms the
conventional dielectric lens for field enhancement. Furthermore, the
device can be mirrored to the other side of the PEC plate as shown
in Fig.~\ref{label:field}(f), and the transmitted electromagnetic
wave will travel through a second distorted space and leads to much
more enhanced field distribution. Fig.~\ref{label:field}(e) shows
that behind the PEC plate the transmitted wave is constrained in a
bundle and high directivity is achieved.

\begin{figure}[t]
\centering
\includegraphics[width=7cm]{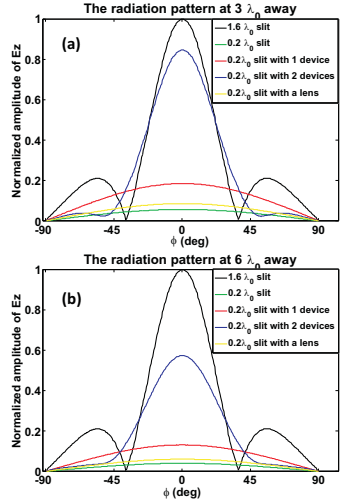}
\caption{(a) The radiation pattern recorded at the semicircle
$3\lambda_{0}$ away from the center of the slit. (b) The radiation
pattern recorded at the semicircle $6\lambda_{0}$ away from the
center of the slit.} \label{label:scattering}
\end{figure}

Radiation patterns are depicted to further compare the field
enhancement performances. In Fig.~\ref{label:scattering}(a), the
amplitude of $E_z$ is recorded along the semicircle of radius
$3\lambda_{0}$ centered in the slit. The $x$ axis is the value of
$\phi$ defined in Fig.~\ref{label:field}(a). The amplitude of the
transmitted $E_z$ field is low when there is only a $0.2\lambda_{0}$
wide slit on the PEC plate. Once the lens is added, the field
increases slightly. However, a significant field enhancement is
observed when the proposed device is put above the slit. The
amplification at $\phi=0\textordmasculine $ is about 3.25 for
amplitude and 10.56 for power. When the mirrored device is added on
the other side, the amplification increases remarkably. Around the
angle of $\phi=0\textordmasculine$, the E field is not much weaker
than that one gets when there is a $1.6\lambda_{0}$ wide slit on the
PEC plate. Note that when the slit is $1.6\lambda_{0}$ wide, the two
side lobes come from the refraction of the PEC plate, which is
already observed in Fig.~\ref{label:field}(b).
Fig.~\ref{label:scattering}(b) shows the radiation pattern at the
semicircle $6\lambda_{0}$ away from the center of the slit (far
field). The patterns in (b) are similar to the patterns in (a). The
amplifications are observed as before while the amplitudes decrease
uniformly when the waves travel further.

\begin{figure}[t]
\centering
\includegraphics[width=7.5cm]{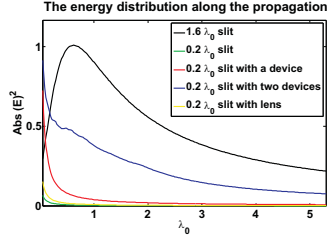}
\caption{The energy distribution along the propagation direction.
The propagating distance is recorded in terms of the wavelength.}
\label{label:propagation}
\end{figure}

Fig.~\ref{label:propagation} also shows the energy distribution
along the propagation path. The energy is recorded at the direction
of $\phi=0\textordmasculine$, from the center of the slit towards
the bottom. When the transmitted waves travel to the far field, the
enhancement of energy is preserved once the designed device is
applied. A pair of devices produce much more transmitted energy than
a single one all along the propagation path. Note that the blue
curve, which represents the energy when a pair of devices are used,
is not smooth from $0.25\lambda_{0}$ to $0.8\lambda_{0}$. This area
is inside the mirrored device and the reflections between different
dielectric blocks are considerable.

\section{The transmitted energy}
\begin{figure}[t]
\centering
\includegraphics[width=6cm]{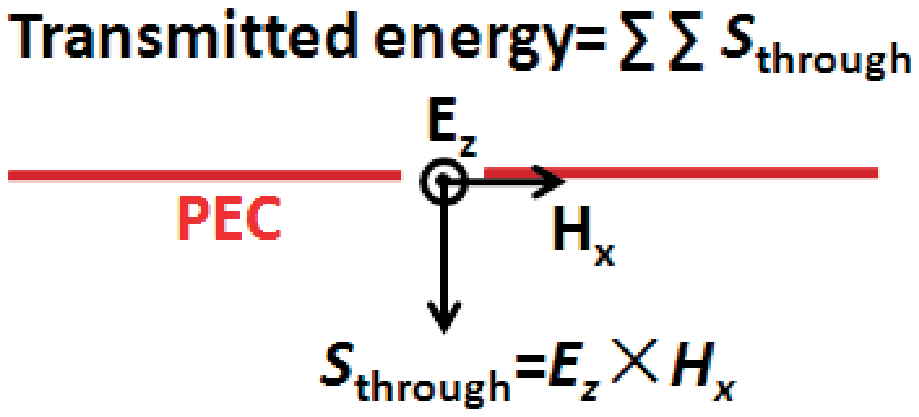}
\caption{Integrate the Poynting vector at the lower surface of the
PEC plate over time to get the transmitted energy.}
\label{label:Poynting}
\end{figure}
To quantify the transmitted energy, a broadband Gaussian pulse is
applied to excite the slit. Its behaviour in the frequency domain is
depicted as the sub-figure in Fig.~\ref{label:Fre1}(a) by using
Fourier transformation. The Poynting vector over the slit represents
the energy passing through. By integrating the Poynting vector at
the lower surface of the PEC plate, the total energy transmitted
over time is available. Notice that at the lower surface of the PEC
plate, only perpendicular electric field and parallel magnetic field
exist. So, the transmitted energy could be calculated as shown in
Fig.~\ref{label:Poynting}. Table 1 gives the values of the
transmitted energy in different cases. It is obvious that extremely
poor transmission is obtained for the 7.5 mm slit if compared with
the energy crossing the 60 mm slit. When a focusing lens is applied,
the enhancement is very limited. But once the enhancement device
studied in this paper is located above the 7.5 mm wide slit, the
transmission increases dramatically (it is about 13 times the energy
transmitted without the device). When applying the second device the
transmission is further increased (the enhancement factor is about
16). With the implementation of the devices, the transmitted energy
increases to about $1/4\sim1/3$ of the value when the slit has an
aperture of 60 mm. The discrepancy between case 1 and case 4 is due
to the fact that the device is mismatched to the free space and the
anisotropic material properties is ignored in the design.

\begin{table}[htb]
\centering \caption{Values of the transmitted energy in different
cases.} \begin{tabular}
{c c l c}\hline
Case No.& & The PEC plate with: & Transmitted energy (arb. units) \\
\hline
1 & &a 60 mm wide slit & 965   \\
2 & &a 7.5 mm wide slit & 19   \\
3 & &a 7.5 mm wide slit and a lens & 28   \\
4 & &a 7.5 mm wide slit and a device & 248  \\
5 & &a 7.5 mm wide slit and a pair of devices & 300
\\ \hline
\end{tabular}
\end{table}

\section{The broadband performance}
\begin{figure}[htbp]
\centering
\includegraphics[width=6.3cm]{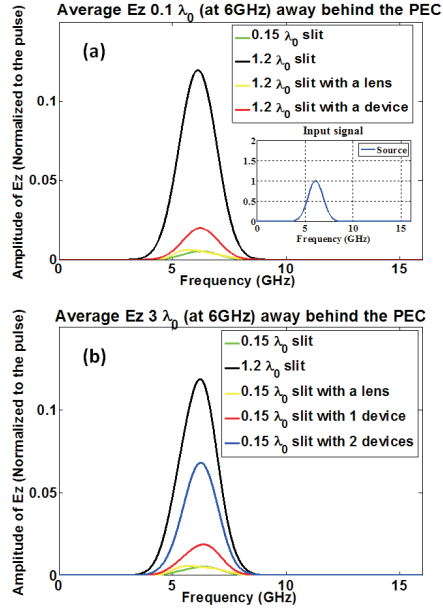}
\caption{(a) The average amplitude of Ez field recorded at the line
$0.1\lambda_{0}$ (at 6 GHz) away behind the PEC plate. (b) The
average amplitude of Ez field recorded at the line $3\lambda_{0}$
(at 6 GHz) away behind the PEC plate.} \label{label:Fre1}
\end{figure}

To investigate the broadband performance of the devices, an incident
pulse spaning from 4 GHz to 8 GHz is applied and the average
amplitude of transmitted Ez field at the lines $0.1\lambda_{0}$ and
$3\lambda_{0}$ (at 6 GHz) away behind the PEC plate is recorded
during every time step. The frequency domain responses are
calculated and plotted in Fig.~\ref{label:Fre1}. Slight enhancement
is observed at some frequencies, when the conventional lens is used.
Once the proposed devices are applied, the transmitted electric
field is considerably higher than the values without any devices
over the whole frequency domain. When the transmitted waves travel
from the very near field of $0.1\lambda_{0}$
[Fig.~\ref{label:Fre1}(a)] to the far field of $3\lambda_{0}$
[Fig.~\ref{label:Fre1}(b)], the significant enhancement is still
held. Note that in Fig.~\ref{label:Fre1}(a), since the observation
line is inside the mirrored device, the case with a couple of
devices is not plotted.

\begin{figure}[t]
\centering
\includegraphics[width=6.3cm]{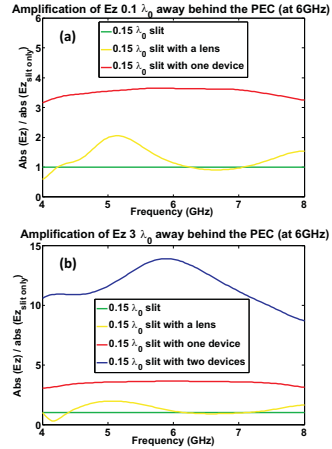}
\caption{(a) Amplifications at the line $0.1\lambda_{0}$ (at 6 GHz)
away behind the PEC plate over the frequency spectrum. (b)
Amplifications at the line $3\lambda_{0}$ (at 6 GHz) away behind the
PEC plate over the frequency spectrum. The amplitude of electric
field is normalized at each frequency point by the value of the
green curve (which represents the transmitted field through the
sub-wavelength slit) to represent the amplification factor.}
\label{label:Fre2}
\end{figure}

The transmission enhancements produced by the devices and the lens
are also calculated and plotted in Fig.~\ref{label:Fre2}. The
amplitude of electric field is normalized at each frequency point by
the value of the green curve, which represents the transmitted field
through the sub-wavelength slit without any devices. The results
give the detailed information of the amplification and prove that
the designed device has significant enhancement performance over a
wide frequency range.

\section{Conclusion}
A device made by dielectric blocks is proposed using the concept of
discrete coordinate transformation. Extraodinary transmission in a
sub-wavelength aperture is achieved over a broad band from 4 GHz to
8 GHz when the device is applied, thanks to the nonresonant nature
of the device. The implementation of a pair of mirrored devices
results in even more transmission and higher directivity behind the
aperture. FDTD based simulation results have verified the expected
performances. The device can be modified to operate for all angles
of incidence by rotating the proposed structures. Overall, the
device demonstrates superior performance over other existing
approaches for broadband extraordinary transmission in a single
slit. More importantly, the design based on transformation
electromagnetics can be applied to optical frequencies as the
required material properties can be easily found from the nature.

\section*{Acknowledgments}
The authors would like to thank the support of the "Spanish Ministry
of Science and Innovation" (Mobility Program, grant number
P2009-0405). They'd also thank Mr. C. Argyropoulos for his help
during the preparation of the manuscript.

\end{document}